\documentstyle[preprint,eqsecnum,aps]{revtex}
\tightenlines

\begin{document}
\draft
\preprint{hep-ph/9803356}
\newcommand{\be}{\begin{equation}}
\newcommand{\ee}{\end{equation}}
\newcommand{\bea}{\begin{eqnarray}}
\newcommand{\eea}{\end{eqnarray}}
\newcommand{\ba}{\begin{array}{l}}
\newcommand{\ea}{\end{array}}
\newcommand{\mufv}{\gamma^{\mu}\gamma^5}
\newcommand{\nufv}{\gamma_\nu\gamma^5}
\newcommand{\gf}{\gamma^5}
\newcommand{\bb}{\begin{thebibliography}}
\newcommand{\eb}{\end{thebibliography}}
\newcommand{\ci}[1]{\cite{#1}}
\newcommand{\lab}[1]{\label{#1}}
\newcommand{\re}[1]{(\ref{#1})}
\newcommand{\Ds}{\displaystyle}
\newcommand{\half}{{\textstyle{\frac{1}{2}}}}
\newcommand{\cM}{{\cal M}}

\title{ On the Role of Charm   in Decay of $B$-mesons to
$\eta '$ $K$ \\}
\author{F. ~Araki, M. ~Musakhanov\footnote
{Associate Member of ICTP,
permanent address: Theoretical  Physics Dept, Tashkent State University,
 Tashkent 700095, Uzbekistan} and H. ~Toki \\}
\address{ Research Center for Nuclear Physics (RCNP), 
Ibaraki, Osaka 567, Japan \\} 
\date{\today}
\maketitle
\begin{abstract}
In this letter we focus on the calculation of $f_{\eta'}^{(c)},$
the amplitude of the $ (c\bar{c}) \to (gluons) \to 
 \eta^{'}$ transition,
which provides the magnitude of the contribution of  the
Cabbibo favored $b\,\rightarrow \,\bar c c s$
elementary process to   $B \to \eta ' K$ decay. 
It is found that $f_{\eta '}^{(c)} \,=\,- 12.3\sim -18.4\, MeV$ 
on the scale of $m_{c}.$
This number is in  strong contradiction with the estimations 
of Halperin, Zhitnitsky and Shuryak, Zhitnitsky
but almost in  agreement with phenomenological analysis of Feldman et al.,
 Petrov
and the estimations of Ali et al.
\end{abstract}
\narrowtext

\section{ Introduction\\ }
\label{Introduction}

Recently there is a great theoretical interest 
\ci{HZ,SZ,AG97,ACGK97,FK97,FK98,F,CT,CT98,DKY,P} 
on the recently released 
experimental data on the branching ratios of the 
decays of $B\,\rightarrow \, K \, \eta '$\ci{CLEO1},
\ci{CLEO2}: 
\begin{eqnarray}
\label{etapkpm}
B( B^\pm \rightarrow \eta^{\prime}  K^\pm) & = & (6.5_{- 1.4}^{+ 1.5}
\pm  0.9) \times 10^{-5},
\\
B(B^0 \to \eta^{\prime} K^0 ) & = & (4.7_{-2.0}^{+2.7} \pm 0.9) \times 
10^{-5}~. 
\end{eqnarray}
In the Standard Model the Cabbibo favored $b\,\rightarrow \,\bar c c s$
elementary process may be followed by conversion of $\bar c c$ pair into
$\eta '$ through the gluons.
The amplitude of this process is described by  
\be
\label{amplitude}
M = \frac{G_{F}}{\sqrt{2}} V_{cb} V_{cs}^* a_1 
< \eta'(p)|\bar c\gamma_{\mu}\gamma_{5}c |0>  < K(q) |
\bar s \gamma_{\mu} b | B(p+q) > .
\ee
Here $G_F$ is the weak coupling constant,  $V_{cb}$, $ V_{cs}^*$ are
Kobayashi-Maskawa matrix elements, $a_{1}=0.25$ phenomenological number
 obtained by a fit (see \ci{HZ} for the references). 
The matrix element 
\be
< \eta ' (p)|\bar c\gamma_{\mu}\gamma_{5}c | 0 >  =
 -i f_{\eta '}^{(c)} p_{\mu}
\lab{fc} 
\ee
is non-zero due to the virtual $ \bar c c \rightarrow  gluons $ transitions. 
Certainly, this matrix element is suppressed by the $ 1/m_{c}^2 $ factor. 
However,
due to strong nonperturbative  gluon fields together with
 the Cabbibo favored  $ b \rightarrow c $ transition 
the suggested mechanism \re{amplitude} can be
expected to  compete appreciably 
with the other mechanisms  of the 
$ B \rightarrow K \eta' $ process \ci{F,DKY}.

If we assume the dominance of the mechanism \re{amplitude}  
   the branching ratio is written in terms of 
 $f_{\eta'}^{(c)}$ as \ci{HZ}
 \be
\label{c}
Br( B \rightarrow K \eta') \simeq 3.92 \cdot 10^{-3} \cdot
\left( \frac{f_{\eta'}^{(c)}}{1 \; GeV} \right)^2 .
\ee
 Using the data \re{etapkpm} it is found 
$f_{\eta'}^{(c)} \simeq 140 \; MeV \; \; (``exp").$ 

This value perfectly coincides with estimation 
of Halperin and Zhitnitsky \ci{HZ}:
\be
\label{HZestimation}
f_{\eta'}^{(c)}  = ( 50 - 180) \; MeV.
\ee
On the other hand,  a recent phenomenological study placed a
bound on $f_{\eta^\prime}^{(c)}$, namely $-65 ~\mbox{MeV} \leq
f_{\eta^\prime}^{(c)} \leq 15  ~\mbox{MeV}$, 
with $f_{\eta '}^{(c)}$ being consistent with zero by analyzing
the $Q^2$ evolution of the  $\eta^\prime \gamma$ form
factor \cite{FK97}, and more recently it was estimated from observed 
ratio of $J/\psi$ decay to $\eta '$ and $\eta_c$ the value of
$f_{\eta '}^{(c)}= -(6.3 \pm 0.6)MeV$ \cite{FK98}. Other similar estimation 
which leads to $|f_{\eta'}^{(c)} | < 12 MeV$  was
made in \ci{P}. 
Ali et al.  considered the complete
amplitude for the exclusive $B-$meson decays, including 
 $\eta^{\prime}  K$ channels, 
where it was combined the contribution from the 
process $b \to s (c\bar{c}) \to s(gluons) \to 
s \eta^{(')}$  with all the others arising from the
four-quark and chromomagnetic operators, as detailed in their papers
\cite{AG97}, \ci{ACGK97}.
Their estimations    gave
 $|f_{\eta^\prime}^{(c)}| \simeq 5.8$ MeV\ci{AG97}
and $f_{\eta^\prime}^{(c)} = -3.1$ ($-2.3$) MeV (for 
$m_c$ in the range $1.3$ - $1.5$ GeV)\ci{ACGK97}  in agreement with the
analysis\cite{FK97}. They stressed  the importance of the 
sign of 
$f_{\eta^\prime}^{(c)}$ and found a theoretical branching ratio
in the range
$$B( B \rightarrow \eta^{\prime}  K) \,=\,(2-4) \times 10^{-5},$$
which is somewhat smaller than the experimental one \re{etapkpm}.
The similar analysis made in \ci{CT} leads the authors to conclude that 
$f_{\eta^\prime}^{(c)}= - 50 MeV$ may provide the explanation of the 
data.
 
Having this situation, it is important to recalculate $f_{\eta^\prime}^{(c)}$ 
to clarify the 
mechanism of $ B \rightarrow K \eta' $ decay in the similar framework performed
by Halperin and Zhitnitsky \ci{HZ}.

\section{Calculations of $\lowercase{f_{\eta^\prime}^{(c)}}$}
\label{Calculations} 
The symmetry of  the classical 
lagrangian may be destroyed by quantum  fluctuations \ci{Schw,Adl,BJa}. 
In gauge theories the axial anomaly arises from noninvariance of 
the fermionic measure against axial transformations 
in the path integrals of the theory\ci{Fujikawa} (see also ref.\ci{FMP},
concerning higher-loop corrections). The present problem is intimately related
with this phenomena.

 In the following we will work only with the Euclidian QCD\footnote
{we are using here the convention  of the Euclidian QCD:\\
$
ix_{M0}=x_{E4},\,\,\,x_{Mi}=x_{Ei},\,\,\,A_{M0}=iA_{E4},\,\,\,
A_{Mi}= - A_{Ei},\,\,\, 
 \psi_{M}=\psi_{E},\,\,\, i\bar\psi_{M}=\psi_{E}^{\dagger},\,\,\,
 \gamma_{M0}=\gamma_{E4} \\ \gamma_{Mi}=i\gamma_{Ei},\,\,\,
 \gamma_{M5}=\gamma_{E5}.
$  We will omit index $E.$}.    
In the Euclidean QCD the axial anomaly in the light quark axial
 current in chiral limit reads
\be
\partial_{\mu}\psi^{\dagger}_{f}\gamma_{5}\gamma_{\mu}\psi_{f} 
= - i  \frac{g^2}{16\pi^2} G\tilde G,
\lab{div}
\ee
where $\psi_{f}$ is the light quark field
$ (f=u,d,s)$ and
 $g$ the QCD coupling constant.  
 $2G\tilde G= \epsilon^{\mu\nu\lambda\sigma}
 G^{a}_{\mu\nu}  G^{a}_{\lambda\sigma}$, where $G^{a}_{\mu\nu}$ is
the gluon field strength operator with $a$ being the color index.

The situation with heavy quarks is very different, since we must
take into account the contribution of the mass term. 
The divergence of the axial current of
charmed quarks has a form:
\be
\partial_{\mu}c^{\dagger}\gamma_{\mu}\gamma_{5}c 
= - i  \frac{g^2}{16\pi^2} G\tilde G    
+ 2m_{c}c^{\dagger}\gamma_{5}c,
\lab{divc}
\ee
The first term in \re{divc} again comes from noninvariance of 
the fermionic measure (or in other words - from Pauli-Villars regularization).
The main problem here is to calculate the contribution from the 
second term in   
\re{divc}. It is clear that this one is reduced to the problem of
 the calculation
of the vacuum expectation value of the operator 
$2m_{c}c^{\dagger}\gamma_{5}c$
 in the  presence of a gluon fields.

In the path integral approach the calculation of the contribution of this term
to any matrix element over light hadrons may be considered in sequence of the 
integrations. Firstly the integration over $c$-quark is performed, 
and the next step is
the calculation of the integral over gauge gluon field and finally integral over
light quarks.

We consider here the first step -- the integration over c-quarks.
We define:
\be
<2m_{c}c^{\dagger}(x)\gamma_{5}c(x)>=
\int DcDc^{\dagger}2m_{c}c^{\dagger}(x)\gamma_{5}c(x)\exp(\int
c^{\dagger}(i\hat\nabla + im_{c})c) .
\lab{def}
\ee
Here $i\hat\nabla \,=\,\gamma_{\mu}(i\partial_{\mu} + g A_{\mu})$
and we introduce the operator $P_{\mu}$ and $p_{\mu}$ 
which are defined in the coordinate space 
as $<x|P_{\mu}|y> \,=\, i\nabla_{\mu}\delta (x-y)$ and 
$<x|p_{\mu}|y> \,=\, i\partial_{\mu}\delta (x-y).$
It is clear that the formal answer for the path integral \re{def} can be
written in the form:
\be
<2m_{c}c^{\dagger}(x)\gamma_{5}c(x)>=
2m_{c}det||\hat P + im_{c}|| <x|Tr\gamma_{5}(\hat P + im_{c})^{-1}|x>
\lab{formal}
\ee
 $det||\hat P + im_{c}||$ must be regularized in the standard manner as 
$$det||\hat P + im_{c}|| \rightarrow 
det||\frac{(\hat P + im_{c})(\hat p + iM)}{(\hat p + im_{c})(\hat P + iM)}||,$$
where $M$ is the regulator mass.
 Eq. \re{formal} must be a gauge invariant function of the gauge field $A$
and therefore must be expressed through  the gluon field strength tensor and
their covariant derivatives.
 
We will follow the operator method  proposed by Vainshtein et al \ci{VZNS83}
in the same line as in \ci{HZ}.
The  key ingredient of this method is based on an assumption 
of a possibility of an
expansion of  \re{formal} over $\frac{gG}{m_{c}^{2}}.$ We will take into account
$O(g^{2}G^2 )$ and $O(g^{3}G^3 )$ terms in the calculations of \re{formal}. We start
from the calculation of 
\begin{eqnarray}
\label{H}
H(x)\,=\,2m_{c} <x|Tr\gamma_{5}(\hat P + im_{c})^{-1}|x> \,=\,\\ 
 -2im_{c}^{2}<x|Tr\gamma_{5}(P^{2}+ m_{c}^{2} + \frac{1}{2}\sigma gG)^{-1}|x>
 \,=\,H_{2}(x)\,+\,H_{3}(x)\,+\,O(g^{4}G^{4}),
\end{eqnarray}
 where
 \be
H_{2}(x) \,=\,-2im_{c}^{2}<x|Tr\gamma_{5}(P^{2}+ m_{c}^{2})^{-2}
  \frac{g}{2} \sigma  G
 (P^{2}+ m_{c}^{2})^{-1} \frac{g}{2}\sigma G |x> 
 \lab{H_2}
 \ee
 and
 \be
H_{3}(x) \,=\,2im_{c}^{2}<x|Tr\gamma_{5}(P^{2}+ m_{c}^{2})^{-2}
\frac{g}{2}\sigma G
 (P^{2}+ m_{c}^{2})^{-1} \frac{g}{2}\sigma G 
 (P^{2}+ m_{c}^{2})^{-1} \frac{g}{2}\sigma G]|x> .
\lab{H^3}
\ee
Here $\sigma G \,=\,\sigma_{\mu\nu}G_{\mu\nu},\,\,\sigma_{\mu\nu}=\frac{i}{2}
[\gamma_{\mu},\gamma_{\nu}].$ 

It is straightforward to calculate $H_{3}(x)$, 
since we may neglect the noncommutativity of the
operators in \re{H^3} and replace $P$ operator by $p$ in \re{H^3}.
In that case we may use the evident formulas,
$$<x|\frac{1}{(p^{2}+ m^{2})^{n}}|x> = \int\frac{d^{4}p}{(2\pi)^{4}}\frac{1}
{(p^{2}+ m^{2})^{n}}=(2^{4}\pi^{2}(n-1)(n-2)m^{2(n-2)})^{-1},$$ 
$$Tr\gamma_{5}(\sigma G)^{3}=i2^{5}tr_{c}G\tilde GG,$$
where $G\tilde GG = G_{\mu\nu}\tilde G_{\nu\rho}G_{\rho\mu}.$
We get 
\be
H_{3}(x) \,=\,-\,i\frac{g^{3}}{2^{4}3\pi^{2} m_{c}^{2}}f_{abc}G^{a}\tilde G^{b}G^{c}
\lab{H^3-1}
\ee

The calculation of  $H_{2}(x)$ needs much more efforts. 
First of all we represent the 
$\sigma G(P^{2}+m_{c}^{2})^{-1}$ in the form
\be
\sigma G  (P^{2}+ m_{c}^{2})^{-1} \,=\, (P^{2}+ m_{c}^{2})^{-1}\sigma G \,+\,
 (P^{2}+ m_{c}^{2})^{-2}[P^{2},\sigma G] \,+\,
 (P^{2}+ m_{c}^{2})^{-3}[P^{2},[P^{2},\sigma G]] \,+\,...
 \lab{expansion}
 \ee
The routine calculations of the commutators in \re{expansion} lead to
\be
[P^{2},\sigma G]\,=\,\nabla^{2}\sigma G \, + \,
 2iP_{\alpha}\nabla_{\alpha}\sigma G ,
\lab{com1}
\ee
\be
[P^{2},[P^{2},\sigma G]]\,=\,\nabla^{4}\sigma G \, + \, 2iP_{\alpha}
(\nabla_{\alpha}\nabla^{2} +  \nabla^{2}\nabla_{\alpha})\sigma G +
(2i)^{2}P_{\beta}G_{\beta\alpha}\nabla_{\alpha}\sigma G 
\,+\,(2i)^{2}P_{\alpha}P_{\beta}\nabla_{\beta}\nabla_{\alpha}\sigma G .  
\lab{com2}
\ee
Other higher commutators lead to the terms order $O(G^{3})$ in the expansion
\re{expansion} and may be neglected. 
Following the arguments of \ci{VZNS83} we may neglect also the terms
which contains single operator $P_{\mu}.$ The reason is that the
matrix elements  
$$<x|(P^{2}+ m_{c}^{2})^{-n}P_{\mu}|x>\sim \nabla_{\mu}G^{2}.$$
By using the Bianchi identity it is easy to show that
\be
\nabla^{2} G_{\mu\nu} = i(G_{\alpha\nu}G_{\alpha\mu}+G_{\alpha\mu}G_{\nu\alpha})
- \nabla_{\nu}\nabla_{\alpha}G_{\alpha\mu} - 
\nabla_{\mu}\nabla_{\alpha}G_{\nu\alpha}
\lab{nabla2G}
\ee
It is evident that the term $\nabla^{4}\sigma G$ is order of $O(G^{3})$ and may be
neglected.
Collecting all of the $O(g^{2}G^{2})$ and $O(g^{3}G^{3})$ terms in  
$H_{2}(x)$ in \re{H_2}, we get
\be
H_{2}(x)\,=\, \frac{i g^{2}}{2^{4}\pi^{2}} G^{a}\tilde G^{a}
\,+\, \frac{i g^{3}}{2^{5}3\pi^{2}m_{c}^{2}} f_{abc}G^{a}\tilde G^{b} G^{c}
 \lab{H^2-1}
 \ee
 Finally
 \be
H(x)\,=\, \frac{i g^{2}}{2^{4}\pi^{2}}G^{a}\tilde G^{a}
 \,-\, \frac{i g^{3}}{2^{5}3\pi^{2}m_{c}^{2}}f_{abc}G^{a}\tilde G^{b} G^{c}
\lab{H-1}
\ee
We neglect here the small contributions of the terms like
$\nabla_{\mu}\nabla_{\alpha}G_{\nu\alpha}.$

As  expected, the first term in $H(x)$ cancels with the first term in 
 \re{divc}, which is the contribution from noninvariance of the measure
and the rest part leads to the divergence of the $c$-quark 
axial current in the form
\be
<\partial_{\mu}c^{\dagger}(x)\gamma_{\mu}\gamma_{5}c(x)> 
\,=\,\,-\, \frac{i g^{3}}{2^{5}3\pi^{2}m_{c}^{2}}f_{abc}G^{a}\tilde G^{b} G^{c} .
\lab{divc-1}
\ee
We would like to stress an attention that our answer for 
$<\partial_{\mu}c^{\dagger}(x)\gamma_{\mu}\gamma_{5}c(x)> $ is $6$ times less
than was calculated by Halperin and Zhitnitsky \ci{HZ}.

We apply this result to the calculation of the $f_{\eta '}^{(c)}.$ The
analogous
quantity $f_{\eta '}^{(u)}$ , which is defined in the similar way 
as $f_{\eta '}^{(c)}$ in\re{fc}, is
\be
m_{\eta '}^{2}f_{\eta '}^{(u)}\,=\,
<0|\frac{g^2}{16\pi^2} G^{a}\tilde G^{a}|\eta ' > .
\lab{fu}
\ee
By definition \re{fc} $f_{\eta '}^{(c)}$ must be calculated from:
\be
m_{\eta '}^{2}f_{\eta '}^{(c)}\,=\,<0| \frac{g^3}{2^{5}3\pi^{2}m_{c}^{2}}
f_{abc}G^{a}\tilde G^{b} G^{c}|\eta ' >
\lab{fc-1}
\ee
The phenomenological way of the estimation of the $f_{\eta '}^{(u)}$ 
is the application of  the 
QCD+QED axial anomaly equation together with data on 
$\eta ' \rightarrow 2 \gamma$
decay leads to 
\be
f_{\eta '}^{(u)} \,=\, 63.6 MeV,
\lab{fu-1}
\ee
which was used in \ci{ACGK97}. 
In the DP chiral quark model \ci{DP}, 
which was successfully checked by the calculation 
of the axial anomaly low-energy theorem \ci{MK} in chiral limit, the 
matrix element in \re{fc-1} \ci{DPW}, \ci{MK} may be reduced to 
the calculation of the matrix element in  \re{fu} with an additional
factor $-\frac{12}{5\rho^{2}}$ \ci{DPW}, \ci{MK}. Here $\rho$ is the  average 
size of the QCD vacuum instantons.
Phenomenological analysis, variational and lattice calculations showed that
\be
\rho\,=\, 1/3 \,fm
\lab{rho}
\ee
So, the ratio of Eqs. \re{fc-1} and \re{fu}
is equal in this model to:
\be
\frac{f_{\eta '}^{(c)}}{f_{\eta '}^{(u)}} \,=\,-\frac{12}{5\rho^{2}}
\frac{1}{6m_{c}^{2}} \,\sim\, -0.1
\lab{ratio}
\ee
By taking into account the estimation \re{fu-1}
(we use $m_c(\mu_1\simeq m_c)\simeq 1.25\,\, GeV$  on the scale 
$\mu_1\simeq m_c$ for the numerical estimates), 
we find
\be
f_{\eta '}^{(c)} \,=\,-\, 6 MeV
\lab{fc-2}
\ee
This number is close to the one of \ci{AG97},
$|f_{\eta '}^{(c)}| \,=\, 5.8 MeV$
and   the sign  and the order of the value coincide with the estimations of 
\ci{ACGK97,FK98,P}.

Recently, Shuryak and Zhitnitsky \ci{SZ} performed direct numerical
evaluations of the various
correlators of the operators $g^{2} G^{a}\tilde G^{a}$, 
$g^{3}f_{abc}G^{a}\tilde G^{b} G^{c}$ in the Interacting Instanton Liquid
Model(IILM).
Their calculations lead to:
\be
<0|g^{2} G^{a}\tilde G^{a}|\eta '> \,=\,7 GeV^3
\lab{GtildeG}
\ee
(which leads to $f_{\eta '}^{(u)} \,=\, 48.3\, MeV$)
and 
\be
\frac{|<0|g^{3}f_{abc}G^{a}\tilde G^{b} G^{c}|\eta ' >|}
{|<0|g^{2} G^{a}\tilde G^{a}|\eta '>|}\approx  (1.5\sim 2.2 )GeV^2
\lab{sz}
\ee
 The later is somewhat large than their simple estimate for this ratio
of matrix elements  
\be 
{12\over 5} < {1\over \rho^2} > \approx  (1\sim 1.5) GeV^2
\lab{est-sz}
\ee
They concluded that 
\be
\lab{SZestimation}
\frac{|f_{\eta '}^{(c)}|}{|f_{\eta '}^{(u)}|}\approx 1.47 \sim 2.11
\ee
On the other hand, with the use of \re{sz} and the abovementioned factor
$-1/6$ in \re{divc-1} we arrive at
\be
\frac{f_{\eta '}^{(c)}}{f_{\eta '}^{(u)}} \,=\,- 0.17\sim - 0.25 .
\lab{ratio1}
\ee
This ratio  gives  $f_{\eta '}^{(c)}\,=\,- 8.2\sim -12.3\, MeV$ at the scale
of the size of the instanton $\mu_2\approx \rho^{-1} .$
The abovementioned experimental numbers \re{etapkpm} are given at
 the scale $\mu_1\approx m_c$, which is
 different from the scale of  this instanton calculation.
 The account of the   anomalous dimension of the 
$g^3G\tilde{G}G$ operator   leads to correction    
\be
\label{correction}
f_{\eta '}^{(c)} (\mu_1\simeq m_c)
\simeq 1.5 f_{\eta '}^{(c)} (\mu_2 \simeq \rho^{-1}) ,
\ee
\ci{SZ}. The account of this scale  factor leads to
\be
\label{fc-4}
f_{\eta '}^{(c)} (\mu_1\simeq m_c)\,=\,- 12.3\sim -18.4\, MeV
\ee
Hence, using \re{sz}, the result of more sophisticated calculations 
of Shuryak and Zhitnitsky, we get
 the number \re{fc-4} which  is 2-3 times 
larger than simple estimation \re{fc-2}.

These numbers \re{fc-2}, \re{fc-4} are in  agreement with the
phenomenological bounds
\cite{FK97,P} and almost in agreement in the sign and the 
value with \ci{ACGK97,FK98} but 
 six-ten times less than the estimations given by  
 \ci{HZ}(see \re{HZestimation}) and also  \ci{SZ} (see \re{SZestimation}).

By using  the numerical analysis of the branching ratio for 
$B^\pm \rightarrow \eta^{\prime}  K^\pm$
 given at
 \ci{AG97} (Fig.17 of \ci{AG97} ) we expect that the value of 
 $f_{\eta '}^{(c)}$ given in  \re{fc-4} may provide a more satisfactory
 explanation of the experimental data\footnote{quite recently Cheng and Tseng
 \ci{CT98} concluded impressive good explanation of these data 
 using  our value \re{fc-4}
for $f_{\eta '}^{(c)}$.}.
 We reserve this investigation for the  future publication.

\acknowledgments
 
 We acknowledge the support of the COE program and partial support by 
 the grant INTAS-96-0597, which enables M.M. to stay at RCNP of Osaka
 University and to perform the present study.

\end{document}